# Accurate Measurement of the Cleavage Energy of Graphite

Wen Wang[1], Shuyang Dai[2], Xide Li[1], Jiarui Yang[1], David J Srolovitz[2,3*], Quanshui Zheng[1,4*]

[1]Department of Engineering Mechanics, Center for Nano and Micro Mechanics, and Applied Mechanics Lab, Tsinghua University, Beijing 100084, China

[2]Department of Materials Science and Engineering, University of Pennsylvania, Philadelphia, PA 19104, USA

[3]Department of Mechanical Engineering and Applied Mechanics, University of Pennsylvania, Philadelphia, PA 19104, USA

[4]State Key Laboratory of Tribology, Tsinghua University, Beijing 100084, China

* e-mail addresses: zhengqs@tsinghua.edu.cn, srol@seas.upenn.edu

**The basal plane cleavage energy (CE) of graphite is a key material parameter for understanding many of the unusual properties of graphite, graphene, and carbon nanotubes. The CE is equal to twice the surface energy and is closely related to the interlayer binding energy and exfoliation energy of graphite. Nonetheless, a wide range of values for these properties have been reported and no consensus has yet emerged as to their magnitude. Here, we report the first direct, accurate experimental measurement of the CE of graphite using a novel method based on the recently discovered self-retraction phenomenon in graphite. The measured value, $0.37 \pm 0.01$ J/m$^2$ for the incommensurate state of bicrystal graphite, is nearly invariant with respect to temperature ($22°C \leq T \leq 198°C$) and bicrystal twist angle, and insensitive to impurities (from the atmosphere). The cleavage energy for the ideal ABAB graphite stacking, $0.39 \pm 0.02$ J/m$^2$, is calculated based upon a combination of the measured CE and a theoretical calculation. These experimental measurements are ideal for use in evaluating the efficacy of competing theoretical approaches.**

Graphite is the most stable form of carbon under standard conditions and is a layered, hexagonal (P6$_3$/mmc) crystal. Each layer is a one-atom thick *graphene* sheet, in which carbon atoms are arranged in a 2D honeycomb lattice (space - plane groups P6/mmc - p6mm) [1-3]. Compared with the extremely strong sp$^2$ intralayer bonds, the interlayer interactions are controlled by much weaker van der Waals bonding. This contrast leads to many novel physical and mechanical properties of graphite, such as maximal values of the electric and thermal conductivities, in-plane elastic stiffness and strength [2,4-9], and the minimum shear-to-tensile stiffness ratio [10]. These novel properties make graphite, graphene, and their allotropes (carbon nanotubes and fullerenes) of intense interest for a wide range of applications.

In spite of a very large and rapidly growing literature on graphite, graphene, and their allotropes, a quantitative understanding and characterization of the interlayer interactions of graphite has yet to emerge [11-20]. The interlayer binding energy is a relatively simple measure of the interlayer interactions and is defined as the energy per layer per area required to separate graphite into individual graphene sheets (e.g., by uniformly expanding the lattice in the direction perpendicular to the basal plane). This energy is nearly equivalent to the exfoliation energy and is approximately equal to the cleavage energy (CE, the energy to separate a crystal into two parts along a basal plane) and twice the basal plane surface energy. On the theoretical side, direct calculation based on conventional density functionals cannot correctly represents the long range van der Waals nature of interlayer interactions in graphite [21]. Recently, several approaches have been suggested to overcome this deficiency, such as Grimme's density functional correction (DFT-D2) [22], a non-local functional (vdW-DF2) [23], the meta-generalized gradient approximation (MGGA-MS2) [24,25], the adiabatic-connection fluctuation-dissipation theorem within the random phase approximation (ACFDT-RPA) [26] and quantum Monte Carlo (QMC) calculations [27,28]. From an experimental perspective, the situation is also murky; there are no reliable, direct measurements of these energies in graphite; previous indirect measurement approaches yield values that range from 0.14 to 0.72 J/m$^2$ (see Table SI of the Supplementary Information for a summary) and no consensus has emerged.

Here, we report the first direct experimental measurement of the cleavage energy (CE) of graphite. The method we adopted is based upon the recently discovered self-retraction phenomenon in graphite [29]. Our experimental method for measuring the CE can be understood in terms of an ideal experiment performed in absolute vacuum as described below. The sample is a rectangular graphite plate adhered to a rigid substrate. The plate itself is a stack of two thinner, single crystal, rectangular graphite flakes, GF1 and GF2, with all (0001) basal planes in both flakes parallel, as illustrated in Fig. 1a. The orientations of the two single crystal flakes are not the same, but are rotated with respect to one another by an angle, $\phi$ ($0 < \phi < 60°$) about the [0001] direction. The interface (grain boundary) energy per unit contact area is denoted $\sigma(\phi)$. The CE of these two flakes is thus $\Gamma_{0001}(\phi) = 2\gamma_{0001} - \sigma(\phi)$, where $\gamma_{0001}$ represents the (0001) surface energy of graphite (of course, at $\phi = 0$, $\sigma = 0$ and $\Gamma_{0001} = 2\gamma_{0001}$). For simplicity, we drop the 0001 subscript since that the remainder paper refers only to the basal plane of graphite.

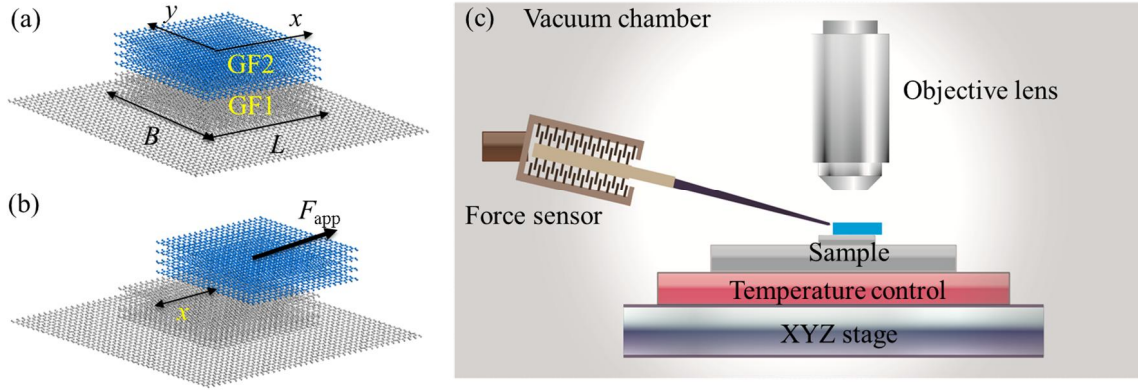

Figure 1. Illustration of the cleavage energy measurement. (a) The graphite sample is a stack of two thin, single crystal, rectangular graphite flakes, GF1 and GF2, with parallel basal planes but rotated with respect to each other about [0001] by an angle, $\phi$. The sample is adhered to a rigid substrate. (b) The cleavage energy is measured through shearing the upper flake relative to the lower one in the superlubric state. (c) Schematic illustration of the experimental setup to shear the sample using an XYZ stage, measure the shear force, $F_{app}$, using a force sensor, and control the temperature and vacuum.

Recent experimental observations showed that the contact between two rotated single crystal basal-oriented graphite flakes is superlubric, namely, the contact is (nearly) frictionless [30-32]. Thus, when slowly shearing the top flake GF1 a distance $x$ (see Fig. 1b), two new free (0001) surfaces with total area $2Bx$ are exposed, where $B$ denotes the flake width. The total free energy changes by $\Delta G = (2\gamma - \sigma(\phi))Bx = \Gamma(\phi)Bx > 0$. As a consequence, a driving force, $F_{ret} = -\frac{dG}{dx} = -\Gamma(\phi)B$ (neglecting any dissipation that may occur – see below), exists for the flake to retract back to its original position (Fig. 1a) in order to reduce the free energy. Therefore, in the superlubric state, the cleavage energy $\Gamma(\phi)$ can be determined through a precise measurement of the applied shear force, $F_{app}$, required to balance the retraction force $F_{ret}$ in the quasi-static loading (shearing) and unloading (retraction) processes: $\Gamma(\phi) = F_{app}/B$. The superlubric retraction process was only recently observed [33].

To perform these experiment, graphite mesas were prepared using the technique reported in [29,31] with the same highly ordered pyrolytic graphite, HOPG (Veeco ZYH and ZYB grade). The HOPG has a brick wall-like polycrystal structure [31] in which each grain is from a few to tens of micrometers wide (parallel to the basal plane) and three orders of magnitude smaller in the perpendicular [0001] direction, ranging from a few to tens of nanometers [31,34]. The grains are stacked such that they share a common [0001] direction but are randomly oriented with respect to that axis. This implies that the grain boundaries perpendicular to [0001] are planer, pure twist boundaries. For our measurements, we prepare mesas with edge lengths $2 \leq B \leq 9$ μm and heights of ~1 μm. Given the dimensions of the grains, mesas frequently have at least one grain boundary parallel to the free surface that runs across the entire mesa [31], as indicated in Figs. 1a,b. These cross-mesa twist grain boundaries are superlubric contacts.

As schematically illustrated in Fig. 1c a micro-force sensing probe (FemtoTools FT-S100 with a 5 nN force resolution and a bandwidth of up to 8 kHz) is fixed to a micro-manipulator (Kleindiek MM3A). The temperature and applied shear force were controlled by placing the test samples on a ceramic heating plate affixed to a stage that can be translated in three dimensions with high precision. In our measurements, the typical rates at which graphite flake GF1 was translated was ~25 nm/s. All of the measurements were performed under an optical microscope (Carl Zeiss Axio Scope.A1).

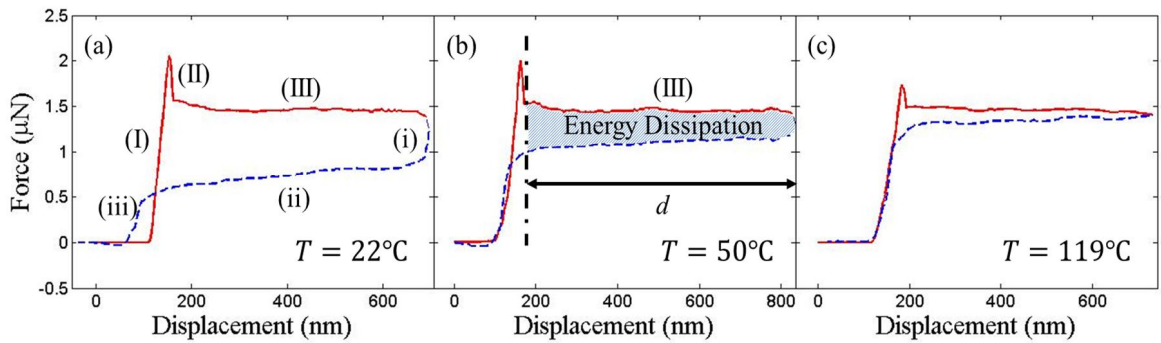

Figure 2 Typical measured force-shear displacement curves for loading (red solid line) and unloading (blue dashed line) in ambient conditions at temperatures (a) 22°C, (b) 50°C, and (c) 119°C. The shaded area in (b) between the entire (III) region loading curve and the unloading curve is the energy dissipated in sliding and retraction. This energy dissipation can be normalized by the (III) region displacement, $d$, and the sample width $B$ to find the dissipative energy (see more below).

We first tested several graphite mesas to verify that self-retraction occurs. For such mesas, we measured the forces and shear displacements of the top flake both during loading and unloading. Fig. 2a shows three typical force-displacement curves for loading and unloading under ambient conditions (temperature $22 \pm 1$°C, relative humidity $25\% \pm 2\%$). The loading curve can be divided into three regions: (I) a nearly linear shear force-displacement region which characterizes the predominantly elastic deformation of the tip before the applied force exceeds the sum of the retraction and static friction force; (II) a sudden drop of the shear force which corresponds to breaking the chemical bonds at the sample edges formed during the reactive ion etch used in fabricating the mesas; and (III) a nearly constant shear force where the applied force $F_{app}$ balances the retraction force $F_{ret}$, $F_{app} = -F_{ret}$ in the superlubric state, where friction is negligible. The slope is zero in loading region III since the advancing flake creates new, contaminant free surfaces as it moves.

Since the loading and unloading cycle required ~100s, the exposed surfaces can absorb a significant quantity of contaminants under ambient conditions [35]. The retracting flake tends to sweep these contaminants [36] ahead of the flake edge in a push broom-like motion that dissipates energy leading to a contamination (or cleaning) friction $F_{cf}$. The unloading curve can also be divided into three regions: (i) an elastic unloading of the tip until $F_{app} \leq -(F_{cf}+F_{ret})$ (recall that $F_{ret} < 0$ and $F_{cf} > 0$); (ii) a region where $F_{cf}$ increases with retraction distance (the advancing flake pushes contaminants ahead of the flake edge – the quantity of contaminant pushed grows in proportion to the flake retraction distance); and (iii) a rapidly decreasing force where GF2 returns to its original position – this overlaps the initial loading region (I) reflecting the elastic unloading of the tip after the upper flake returns to its initial position.

To validate the conjectured role of impurities in creating contamination friction $F_{cf}$, we performed similar loading and unloading measurements as a function of temperature in the same environment – see Figs. 2a-c. The expectation is that increasing temperature reduces the equilibrium impurity concentration on the newly exposed surfaces [35]. Examination of Fig. 2 shows that the gap between the loading and unloading curves and the slope on the unloading (retraction) curve (region (ii)) decreases with increasing temperature. The decrease in the slope in the $F_{app}$ versus displacement curve in unloading (region (ii)) with increasing temperature is associated with decreased impurity concentration on the surface at higher temperature; recall that $F_{cf}$ is proportional to the area of the surface swept (sliding distance) during translation of the upper crystal with respect to the lower one during retraction. Hence, $F_{cf}$ should go to zero in the high temperature limit (see Fig. 3a inset); the temperature at which this term becomes negligible should scale in proportion to the contaminant – surface binding energy. The fact that the loading and unloading curves are nearly identical at the highest temperature (119°C) demonstrates that there is little hysteresis in the sliding/retraction process. Additional results over a wider temperature range are shown in Fig. 3a. Additionally, the fact that the loading curve in region (III) is nearly identical to the unloading (retraction) curve in region (ii) at slightly elevated temperatures (see Fig. 2c) demonstrates that the magnitude of the dynamic friction force is negligible (since this force points in opposite directions on loading and unloading) and the retraction (above ~100°C) is superlubric. Finally, we note that since the loading curve is flat and temperature-independent, the flake translation on loading is superlubric over the entire temperature range examined.

These observations, taken together, clearly demonstrate that $F_{\text{app}} = -F_{\text{ret}} = B\Gamma(\phi)$ or that measurements of $F_{\text{app}}$ (in region III) and the sample width ($B$) give the cleavage energy, $\Gamma(\phi) = F_{\text{app}}/B$. In this manner, we find an average cleavage energy of **$\Gamma(\phi) = 0.37 \pm 0.01$** J/m$^2$, where the data was averaged over 50 samples with 2-9 µm flakes with rotation angles $16° \leq \phi \leq 54°$.

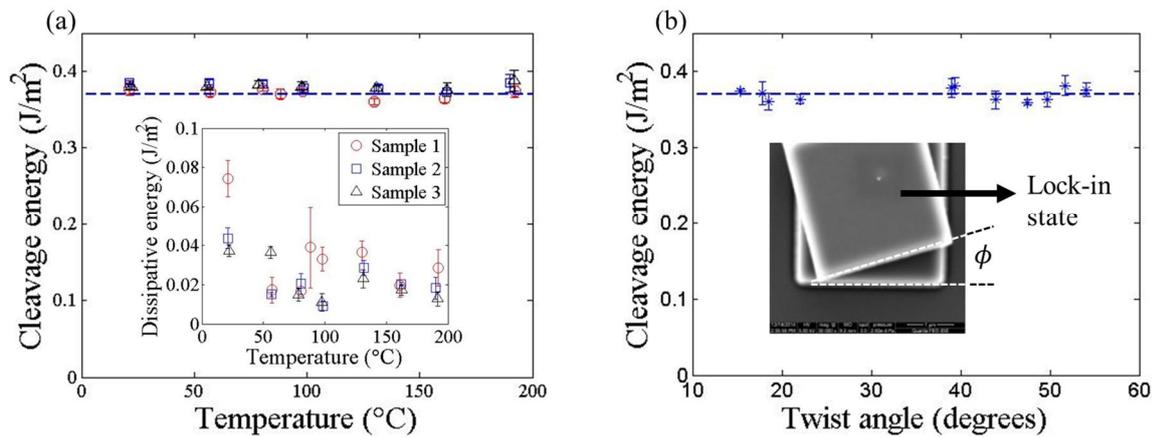

Figure 3 (a) The measured cleavage energy as a function of temperature. Red, blue and black symbols represent 3 different samples with the same side length ($B = 4$ µm) and the error bars represent the standard deviation of 5 independent measurements on each. The inset shows the dissipative energy, defined in the caption of Figure 1(b), versus temperatures. (b) The measured cleavage energy as a function of twist angle $\phi$ at T=20°C. The CE values were measured from 11 different samples with the same side length ($B = 3$ µm). Each sample corresponds to a different twist angle. The error bars represent the standard deviation of 5 independent measurements for each sample. The insert shows a typical sample at lock-in from which the rotation angle is measured.

At finite temperature, we expect the individual basal planes to fluctuate. This could give rise to a thermal effect on the CE; such an effect has not heretofore been reported. We experimentally investigate the impact of temperature in ambient laboratory conditions (at a relative humidity of 22% ± 5% and with different temperatures from 22°C to 198°C). Fig. 3a shows the measured CE as a function of temperature (based upon the loading curves). From these results, we see that the CE of incommensurate (large twist angle) graphite is nearly temperature invariant over the temperature range examined. On the other hand, the shaded area between regions (III) on loading and region (ii) (see Fig. 2) is clearly temperature dependent. Normalizing by the displacement $d$ and the sample length $B$ gives an intrinsic measure of this effect. The insert in Fig. 3a shows that the dissipative energy decreases with increasing temperature. As discussed above, this is likely due to decreased contaminant concentration on the surface with increasing temperature. We have not examined whether this represents the equilibrium adsorption isotherm or kinetics plays a role here.

As discussed above, the cleavage energy of graphite $\Gamma$ is the difference between twice the surface energy $2\gamma$ and the twist grain boundary energy $\sigma$. Since $\sigma$ is expected to be a function of twist angle $\phi$ (like grain boundaries in most crystalline materials), so too is $\Gamma$. The first step in determining this $\phi$-dependence is the measurement of $\phi$. We do this based upon the lock-in effect [31]; this refers to the observation that self-retraction disappears at a particular rotation angles of GF2 relative to GF1 [31]. This can be understood as follows: if two crystals have an arbitrary rotation with respect to one another such that they are incommensurate and the two crystals are rigid, there is no barrier to sliding [37-39]. However, when two graphite flakes are commensurate (perfect ABAB stacking) at $\phi = 0$, the barrier to sliding is the theoretical shear strength of the material. This was observed in Ref. [40]. By measuring the angle required to rotate GF2 into such a no-retraction condition, we determine the initial rotation of GF2 relative to GF1, i.e, $\phi$. Fig. 3b shows the cleavage energy as a function of twist angle $\phi$ (the inset shows a typical observation of a flake rotated into the no-retraction condition). These results, obtained from 11 samples of the same side length $B = 3$ µm, show that over the range of angles examined ($16° \leq \phi \leq 54°$), the cleavage energy is surprisingly independent of twist angle $\phi$.

While several measurements and predictions (see Tables S1 and S2 in the Supplementary Information) are available for interlayer bonding and the surface energy of graphite ($\gamma = \Gamma(0)/2$), little information is available on the twist boundary energy $\sigma(\phi)$. We turn to theoretical analysis to understand both the magnitude of $\sigma$ and its independence on twist angle, $\phi$. We do this in the framework of the Peierls-Nabarro model [41-43] (that was originally formulated to describe dislocations), generalized to account for anisotropic elasticity [44] and extended to describe twist boundaries [45,46]. In this model, the total energy consists of two parts: the elastic energy stored in the crystals on either side of the boundary and the misfit energy that represents the atomic interactions (bonding) between the two crystals (at the grain boundary). The only inputs to the model are the anisotropic elastic constants for graphite and the generalized stacking fault energy (GSFE). The GSFE is simply the total energy of a pair of semi-infinite rigid graphite crystals meeting at a (0001) plane as a function of the shift of the two crystals parallel to that plane minus the energy when the shifts are zero (i.e., perfect ABAB stacking). The form of the two dimensional GSFE function (displacements in two orthogonal directions in the (0001) plane) must respect the symmetry of the graphite crystal structure [45,46]. In the Supplementary Information, we describe how the GSFE is obtained based upon accurate first-principles calculations and provide all of the functions and parameters used as input to the anisotropic Peierls-Nabarro grain boundary (APNGB) model applied to (0001) twist grain boundaries in graphite.

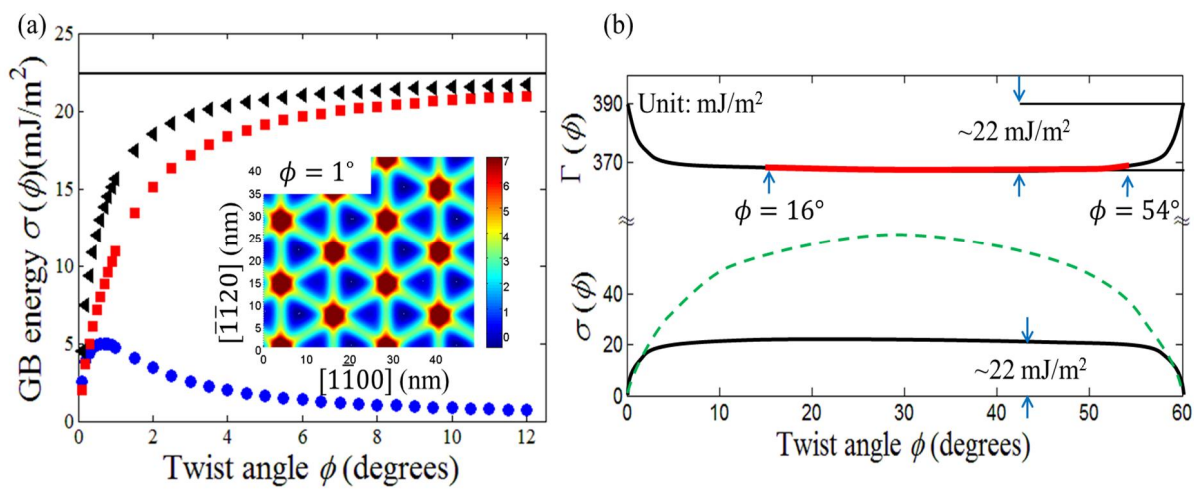

Figure 4 (a) The (0001) twist grain boundary (GB) energy in graphite obtained from the APNGB model as a function of twist angle $\phi$ (black symbols). The contributions to the GB energy from the elastic and misfit (GSFE) energies are shown by blue and red symbols. The solid black horizontal line shows the GB energy assuming that the graphite crystals are rigid and incommensurate with each other (obtained as the average over the entire GSFE). The inset shows the spatial distribution of the misfit energy in a $\phi=1°$ twist grain boundary (dislocations cores are seen as green lines). (b) The results from (a) plotted over the entire twist angle range (lower black curve) and the corresponding cleavage energy $\Gamma(\phi)$ over the same range based upon the measured values of $\Gamma(\phi)$ (shown as the red line). The green dashed curve shown for comparison, is the typical shape of $\sigma(\phi)$ for a twist GB on the (111) plane of a face centered cubic metal [46].

For small, twist angles $\phi$, the grain boundary can be thought of as a two dimensional array of dislocations [43,45], as shown in the inset to Fig. 4 a for a 1° twist angle from the anisotropic Peierls-Nabarro grain boundary (APNGB) calculation. The green lines (intermediate misfit energy) in that figure represent the dislocation cores and the red regions with highest misfit energy are the positions where dislocation lines intersect. These results show that the dislocation core width is $w \simeq 3$ nm (the width of the green lines in the inset to Fig. 4a). This exceptionally large dislocation core width is associated with the very weak interlayer bonding and relatively strong/stiff intralayer bonding in graphite and is consistent with electron microscopy observations [47]. The nearly triangular dislocation array is associated with the dissociation of the screw dislocations in this boundary into partial dislocations and alternating triangles correspond to regions of ABAB|ABAB (perfect crystal) stacking and ABAB|CACA stacking (i.e., a stacking fault). The magnitude of the stacking fault energy in graphite is very small, ~0.85 mJ/m² (see the Supplementary Information). The corresponding twist boundary energy versus twist angle is shown in Fig. 4a. The energy rises rapidly from zero at 0° and saturates at ~22 mJ/m$^2$ over a characteristic angle range of 4° (90% of the saturation value). The saturation of the twist boundary energy at such a small angle is unusual compared with non-van der Waals bonded materials (e.g., metals [45,46]) and can be understood as the angle for which the dislocation cores overlap. The dislocation spacing is $d \simeq b/\phi$, where $b$ is the magnitude of the Burgers vector (~0.2 nm for partial dislocations in graphite). Hence, the critical angle for dislocation overlap, i.e., $d \simeq w$, is $\phi_c \simeq \frac{b}{w} \simeq \mathbf{3.7°}$, in good agreement with the APNGB results. A similar condition applies at 60°–$\phi_c$, where the 60° rotation corresponds to a perfect twin with extremely small energy. For twist angles in the range $\phi_c \leq \phi \leq \mathbf{60°} - \phi_c$, the dislocation cores significantly overlap and the twist boundary can be viewed as two rigid crystals meeting incommensurately at the twist boundary. The energy of such a configuration is almost entirely the result of the misfit energy (the elastic energy is negligible over this angle range – see Fig. 3 a) and can be simply obtained by performing an average over the entire generalized stacking fault energy (see the Supplementary Information). This is the asymptotic, large angle grain boundary energy, which is $\sigma_0 \simeq \mathbf{22}$ mJ/m$^2$ for graphite.

These theoretical results can be used to interpret the experimental findings. The cleavage energy is predicted to be nearly independent of twist angle over the entire experimental range from 16° to 54°. This is consistent with the experimental observations (Fig. 3b). The theoretical results show that a variation with twist angle should only be seen for $\mathbf{0°} < \phi < \mathbf{4°}$ or $\mathbf{56°} > \phi > \mathbf{60°}$. Since the contribution to the cleavage energy from the surface energy is so much larger than the grain boundary energy (and its variation), even for these angles, the variation in $\Gamma$ with $\phi$ will be small. We can use the theoretical results to estimate the (0001) surface energy from the experimentally measured cleavage energies. Over the experimentally accessible twist angle range, with a measured value of $\Gamma = \mathbf{0.37 \pm 0.01}$ J/m$^2$, the ideal cleavage energy is $\Gamma(\mathbf{0}) = \Gamma + \sigma_0 = 2\gamma = \mathbf{0.37 + 0.02}$ J/m$^2 = \mathbf{0.39 \pm 0.02}$ J/m$^2$, where $\sigma_0$ is the large angle ($\mathbf{4°} \leq \phi \leq \mathbf{56°}$) value of the twist grain boundary energy. We estimate the error in $\sigma_0$ to be less than ~0.005 J/m$^2$ (see the Supplementary Information).

Graphite is an unusual material; it has very strong (covalent) bonding within the basal plane but has extremely weak (van der Waals) bonding between basal planes. This results in very large (small) values of the elastic constants with components without (with) components in the direction normal to the basal plane. While unusual compared with most materials, it is also prototypical of layered van der Waals bonded systems. Hence, definitive values for the main energetics of this system are both interesting and important. This has led to a wide range of measurements and theoretical predictions of the strength of this bonding (especially as it relates to the interlayer bonding). This interlayer binding energy has been reported in several forms for graphite [48]; namely the cleavage energy (CE), the (0001) surface energy (SE), the binding energy (BE), and the exfoliation energy (EE, the energy per area required to remove one (0001) atomic layer from the surface of the bulk material). Experimental measurements suggest cleavage energies in the 0.19-0.72 J/m$^2$ range (or $0.43 \pm 0.29$ J/m$^2$) and the theoretical predictions are in the 0.03-0.51 J/m$^2$ range (or $0.27 \pm 0.24$ J/m$^2$); see Tables S1 and S2 in the Supplementary Information. Prior to the present work, direct, accurate experimental measurements of these energies were unavailable and theoretical predictions were routinely confounded by the difficulty of fully including dispersion forces within first-principles frameworks (even the most accurate methods show significant variations). The different measurements of the interlayer bonding are inter-related by either exact relations or by theoretical estimates; SE = CE/2, BE $\approx$ 0.85 CE and CE $\approx$ 0.85 EE [48]. In the present work, we report accurate experimental results for the CE of incommensurate graphite on the basal plane, i.e., CE = $0.37 \pm 0.01$ J/m$^2$. In order to relate this to the CE of a perfectly stacked AB graphite crystal, we performed anisotropic Peierls-Nabarro grain boundary energy calculations based on a combination of experimental and first principle results to obtain a grain boundary energy with a maximum value of $\sigma = \mathbf{0.02 \pm 0.005}$ J/m. This implies CE = $0.39 \pm 0.02$ J/m$^2$ for perfectly stacked AB graphite. While this value is a combination of experimental and computational results, the uncertainties are still very small and this value should be considered definitive. Using the relations described above, these results imply a basal plane surface energy of 0.20 J/m$^2$, an interlayer binding energy of 0.33 J/m$^2$ and exfoliation energy of 0.46 J/m$^2$. These results provide an excellent means to distinguish between competing approaches for *ab initio* prediction of bonding in van der Waals materials.

## Acknowledgements

The experimental was performed with the support of NSFC (Grant No. 11227202), the National Basic Research Program of China (Grant Nos. 2013CB934203 and 2010CB631005), and SRFDP (Grant No. 20130002110043). The theoretical work was supported by the Office of Sciences, Basic Energy Sciences, US Department of Energy, EFRC award DE-SC0012575.